\pdfoutput=1
\documentclass[prl, reprint, aps]{revtex4-1}
\usepackage{graphicx}
\usepackage{hyperref}

\usepackage{xspace}

\newcommand{\ca}{$\mathrm{^{40}Ca^{+}}$\xspace}
\newcommand{\tran}[2]{#1 $\leftrightarrow$ #2\xspace}
\newcommand{\qs}[2]{$\mathrm{#1_{#2}}$}

\newcommand{\SI}[2]{\ifmmode #1\enskip\mathrm{#2}\else
$#1\enskip\mathrm{#2}$\fi} 
\newcommand{\wz}[1]{\ifmmode \nu_z = \SI{#1}{kHz} \else $\nu_z=\SI{#1}{kHz}$}

\newcommand{\nuhigh}{\wz{583}\xspace}
\newcommand{\nulow}{\wz{75}\xspace}

\newcommand{\mJ}[1]{$m_J=#1/2$}

\newcommand{\mn}{\ifmmode \langle n \rangle\else $\langle n \rangle$\xspace\fi}

\newcommand{\wzM}[1]{\ifmmode \nu_z = \SI{#1}{MHz} \else $\nu_z=\SI{#1}{MHz}$}

\begin{document}
\date{\today}
\title{Adiabatic cooling of a single trapped ion}
\author{G. Poulsen}
\author{M. Drewsen}
\affiliation{QUANTOP - The Danish National Research Foundation Center for Quantum Optics\\ Department for Physics and Astronomy, Aarhus University, Denmark}

\begin{abstract}

We present experimental results on adiabatic cooling of a single \ca ion in a
linear radiofrequency trap. After a period of laser cooling, the secular
frequency along the rf-field-free axis is adiabatically lowered by nearly a
factor of eight from \nuhigh to \nulow. For an ion originally Doppler laser
cooled to a temperature of \SI{0.65 \pm 0.03}{mK}, a temperature of \SI{87 \pm
7}{\mu K} is measured after the adiabatic expansion. Applying the same adiabatic
cooling procedure to a single sideband cooled ion in the ground state ($P_0 =
0.978\pm0.002$) resulted in a final ground state occupation of $0.947\pm0.005$.
Both results are in excellent agreement with an essentially fully adiabatic
behavior.  The results have a wide range of perspectives within such diverse
fields as ion based quantum information science, high resolution molecular ion
spectroscopy and ion chemistry at ultra-low temperatures.
\end{abstract}

\maketitle

Cooling through adiabatic expansion is a general and well-known method to reduce
the temperature of a gaseous system. For instance, within cold atomic gas
physics, adiabatic cooling has previously been demonstrated with neutral atoms
by applying optical lattices \cite{PhysRevLett.69.1344, PhysRevLett.74.1542}.
Here, one of the most spectacular results has been the possibility to study the
transition from the MOT insulating to the superconducting phase of ultracold
atoms \cite{Greiner2002}. Adiabatic cooling of trapped ions was discussed severeal
decades ago \cite{Bergquist1988,springerlink:10.1007/BF01438559}, while only recently
adiabatic cooling played a major role in reaching the lowest temperatures for
ensembles of antiprotons ever \cite{PhysRevLett.106.073002}. So far, pure
adiabatic cooling of a single trapped ion has not yet been investigated
experimentally.

For ion based quantum information science, adiabatic motional dynamics has so
far mainly been a concern in connection with experiments on shuttling
\cite{PhysRevLett.102.153002,PhysRevA.84.032314} and separating of ions
\cite{Barrett2004}, but for ion quantum logic gates based on magnetic field
gradients \cite{PhysRevLett.87.257904, PhysRevLett.101.090502}, separation of
ions through adiabatic opening of the trap potential could be an advantage.
Furthermore, adiabatic cooling could possibly become a vital tool for both high
resolution molecular ion spectroscopy \cite{schmidt:305, Vogelius:Probabilistic,
PhysRevLett.98.180801, Rosenband28032008, PhysRevA.71.032505,
PhysRevLett.99.150801} and ultracold ion chemistry \cite{B813408C,
1367-2630-11-5-055049}. With respect to the first topic, the possibility to
reach low ion oscillation frequencies while staying in the motional ground
state may open up for quantum logic spectroscopy \cite{Schmidt29072005} on
rather long wavelength vibrational transitions; e.g., for testing fundamental
physics \cite{PhysRevLett.98.180801, Rosenband28032008, PhysRevA.71.032505,
PhysRevLett.99.150801}. For the latter topic, adiabatic cooling could make it
possible to enter the ultracold regime ($\mathrm{\mu K}$ and below) where new
chemistry is expected \cite{PhysRevLett.89.093001, PhysRevA.79.010702}.

In this paper, we present a detailed investigation of adiabatic cooling a single
\ca ion in a linear rf quadrupole trap. Through adiabatic lowering of the
secular frequency along the rf free axis from a value of \nuhigh down to \nulow,
we demonstrate cooling from Doppler cooled conditions (\SI{T = 0.69\pm0.02}{mK})
down to \SI{T=87 \pm 7}{\mu K} - equivalent to a temperature six times lower
than the theoretic Doppler limit (\SI{T = 0.5}{mK}). For the same adiabatic
procedure, we furthermore  show that an ion sideband cooled to the ground state
($P_0 = 0.978\pm0.002$) can be kept in this state with a probability above 97\%.

The setup and procedures used are the same as described in Ref.
\cite{Poulsen:SBC} except for the sideband cooling step. In the current work,
sideband cooling is realized by a continuous scheme where the ion is addressed
simultaneously by light at 729 nm and 854 nm. The light at 729nm excites the ion
on the red sidebands of the \tran{\qs{S}{1/2}(\mJ{-1})}{\qs{D}{5/2}(\mJ{-5})}
transition while the 854nm light acts to broaden the upper level by coupling it
to the ground state via excitation to the higher lying \qs{P}{3/2} level. The
sideband cooling starts with 0.5 ms cooling on the 2nd red sideband followed by
0.5 ms cooling on the first red sideband with optical power corresponding to
Rabi frequencies in the range of few hundreds of kHz. This power provides fast
cooling but at the cost of significant off resonant excitation. To reach the
ground state with high probability, the cooling sequence is finalized by a low
power step with a Rabi frequency around 30 kHz lasting 5 ms. During all cooling
steps, the 729nm light is additionally switched to the
\tran{\qs{S}{1/2}(\mJ{+1})}{\qs{D}{5/2}(\mJ{-1})} transition every \SI{100}{\mu
s} to avoid trapping in the \qs{S}{1/2}(\mJ{+1}) state.
\begin{figure*}
\includegraphics{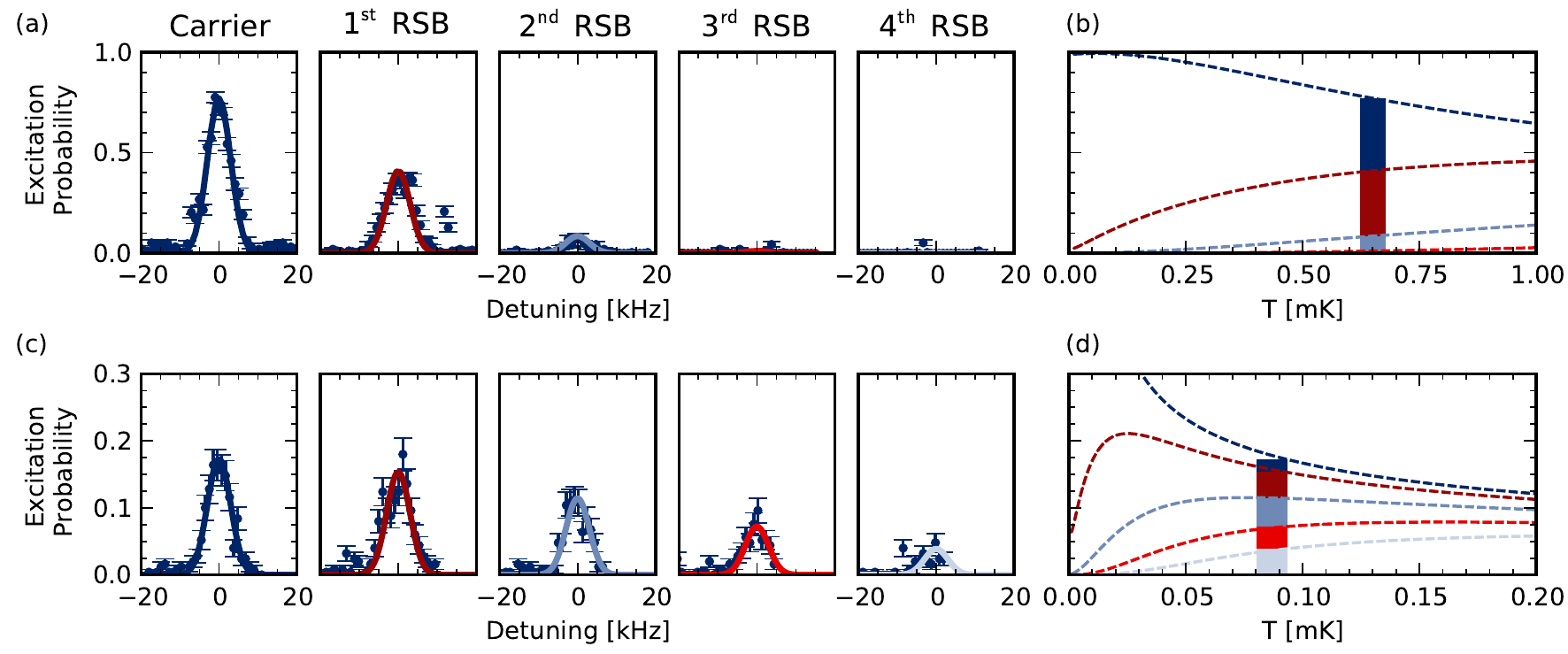}

\caption{Sideband spectra before and after adiabatic lowering of the secular
frequency along the rf field free axis. (a)  Excitation probabilities for the
carrier and the first four sidebands after Doppler cooling including best fits
for a thermal motional distribution. The horizontal axis indicates the detunings
 with respect to the sideband frequencies $n\cdot\nu_z$ (\nuhigh). (b) Expected
 sideband excitation probabilities versus temperature for a thermal
motional distribution. The vertical bars indicates the height of the various
sidebands and the width the standard deviation extracted from the fit. The
vertical axis is the same as in (a). (c) As (a) but after adiabatically opening
the trap potential (\nulow). (d) As (b) after lowering secular frequency
(\nulow).}
\label{fig:doppler}
\end{figure*}

After a period with laser cooling (5 ms Doppler cooling optionally followed by 6
ms of sideband cooling) at the high secular frequency, the laser light is
switched off and the axially confining dc potential applied to the eight
end-electrodes is lowered through a filtering circuit (used to combine the rf
and dc potentials) with a characteristic time constant of 1 ms. A 5 ms delay is
inserted before starting the measurements of the motional state of the ion to
assure the axial trap potential has settled. The rather long ramp time safely
situates the experiments at all time within the adiabatic regime set by
$\dot{\nu_z}/\nu_z^2\ll2\pi$ , where $\nu_z$ is the relevant secular frequency.
At the few millisecond time scale, serious problems concerning spurious heating
of the ion is not expected since we have consistently measured heating rates of
one vibrational quanta per second in the frequency range 280 - 585 kHz
\cite{Poulsen:SBC}.


For the Doppler cooling experiments, the initial and final temperature is
derived by comparing the excitation spectrum spanning the carrier and four
lowest order red sidebands with those expected assuming a thermal distribution.
In \autoref{fig:doppler}, such spectra are presented both before (a), and after
(c), the adiabatic cooling. To extract the temperature, the excitation
probabilities for the sidebands are matched to a thermal model as illustrated in
(b) and (c). Clearly, the excitation matches the model well and suggests initial
and final temperatures of \SI{0.65 \pm 0.03}{mK} and \SI{87 \pm 7}{\mu K}
respectively. The ratio of these temperatures match that of the secular
frequencies, 7.8, within the uncertainties indicating good adiabatic behavior.

\begin{figure}
\includegraphics{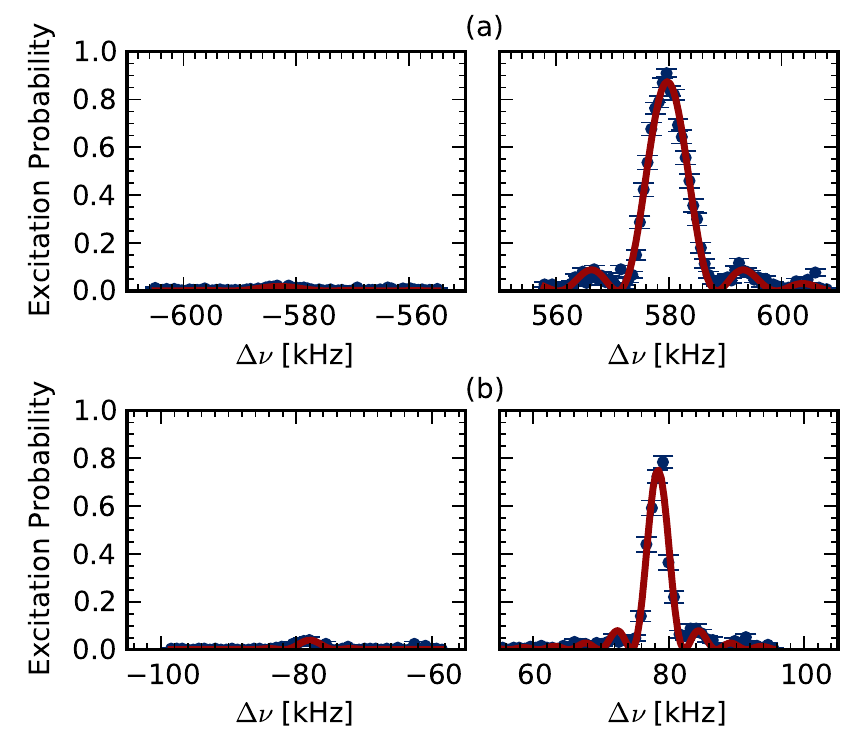}

\caption{Comparison of red and blue sidebands before (a) and after (b) adiabatic
lowering the secular frequency. The spectra indicate ground state populations of
$0.978\pm0.02$ and $0.947\pm0.005$ respectively.}
\label{fig:sbc}
\end{figure}

In the sideband cooled case, the ground state population is deduced by comparing
the excitation probabilities when addressing the first red and blue sidebands
\cite{PhysRevLett.83.4713}. In \autoref{fig:sbc}, these spectra are presented
(a) before,  and (b) after, the adiabatic cooling process. \autoref{fig:sbc}a
suggests a ground state population of $97.8\pm0.2$\%, while from
\autoref{fig:sbc}b, a value of $94.7\pm0.5$\% can be inferred. Again, a
near perfect adiabatic cooling has been achieved, though with indication of
minor heating. The final state distribution
corresponds to a translational temperature of $\sim\SI{1.3}{\mu K}$.

In conclusion, we have demonstrated very effective adiabatic cooling of an ion
cooled either initially to near the Doppler limit or to its ground state of motion.
In both cases, essentially perfect adiabatic cooling has been achieved.
The demonstrated adiabatic cooling will likely find applications within a wide
range of ion based research, including quantum information science, high
resolution molecular ion spectroscopy and ultracold ion chemistry.


\begin{thebibliography}{24}%
\makeatletter
\providecommand \@ifxundefined [1]{%
 \@ifx{#1\undefined}
}%
\providecommand \@ifnum [1]{%
 \ifnum #1\expandafter \@firstoftwo
 \else \expandafter \@secondoftwo
 \fi
}%
\providecommand \@ifx [1]{%
 \ifx #1\expandafter \@firstoftwo
 \else \expandafter \@secondoftwo
 \fi
}%
\providecommand \natexlab [1]{#1}%
\providecommand \enquote  [1]{``#1''}%
\providecommand \bibnamefont  [1]{#1}%
\providecommand \bibfnamefont [1]{#1}%
\providecommand \citenamefont [1]{#1}%
\providecommand \href@noop [0]{\@secondoftwo}%
\providecommand \href [0]{\begingroup \@sanitize@url \@href}%
\providecommand \@href[1]{\@@startlink{#1}\@@href}%
\providecommand \@@href[1]{\endgroup#1\@@endlink}%
\providecommand \@sanitize@url [0]{\catcode `\\12\catcode `\$12\catcode
  `\&12\catcode `\#12\catcode `\^12\catcode `\_12\catcode `\%12\relax}%
\providecommand \@@startlink[1]{}%
\providecommand \@@endlink[0]{}%
\providecommand \url  [0]{\begingroup\@sanitize@url \@url }%
\providecommand \@url [1]{\endgroup\@href {#1}{\urlprefix }}%
\providecommand \urlprefix  [0]{URL }%
\providecommand \Eprint [0]{\href }%
\providecommand \doibase [0]{http://dx.doi.org/}%
\providecommand \selectlanguage [0]{\@gobble}%
\providecommand \bibinfo  [0]{\@secondoftwo}%
\providecommand \bibfield  [0]{\@secondoftwo}%
\providecommand \translation [1]{[#1]}%
\providecommand \BibitemOpen [0]{}%
\providecommand \bibitemStop [0]{}%
\providecommand \bibitemNoStop [0]{.\EOS\space}%
\providecommand \EOS [0]{\spacefactor3000\relax}%
\providecommand \BibitemShut  [1]{\csname bibitem#1\endcsname}%
\let\auto@bib@innerbib\@empty
\bibitem [{\citenamefont {Chen}\ \emph {et~al.}(1992)\citenamefont {Chen},
  \citenamefont {Story}, \citenamefont {Tollett},\ and\ \citenamefont
  {Hulet}}]{PhysRevLett.69.1344}%
  \BibitemOpen
  \bibfield  {author} {\bibinfo {author} {\bibfnamefont {J.}~\bibnamefont
  {Chen}}, \bibinfo {author} {\bibfnamefont {J.~G.}\ \bibnamefont {Story}},
  \bibinfo {author} {\bibfnamefont {J.~J.}\ \bibnamefont {Tollett}}, \ and\
  \bibinfo {author} {\bibfnamefont {R.~G.}\ \bibnamefont {Hulet}},\ }\href
  {\doibase 10.1103/PhysRevLett.69.1344} {\bibfield  {journal} {\bibinfo
  {journal} {Phys. Rev. Lett.}\ }\textbf {\bibinfo {volume} {69}},\ \bibinfo
  {pages} {1344} (\bibinfo {year} {1992})}\BibitemShut {NoStop}%
\bibitem [{\citenamefont {Kastberg}\ \emph {et~al.}(1995)\citenamefont
  {Kastberg}, \citenamefont {Phillips}, \citenamefont {Rolston}, \citenamefont
  {Spreeuw},\ and\ \citenamefont {Jessen}}]{PhysRevLett.74.1542}%
  \BibitemOpen
  \bibfield  {author} {\bibinfo {author} {\bibfnamefont {A.}~\bibnamefont
  {Kastberg}}, \bibinfo {author} {\bibfnamefont {W.~D.}\ \bibnamefont
  {Phillips}}, \bibinfo {author} {\bibfnamefont {S.~L.}\ \bibnamefont
  {Rolston}}, \bibinfo {author} {\bibfnamefont {R.~J.~C.}\ \bibnamefont
  {Spreeuw}}, \ and\ \bibinfo {author} {\bibfnamefont {P.~S.}\ \bibnamefont
  {Jessen}},\ }\href {\doibase 10.1103/PhysRevLett.74.1542} {\bibfield
  {journal} {\bibinfo  {journal} {Phys. Rev. Lett.}\ }\textbf {\bibinfo
  {volume} {74}},\ \bibinfo {pages} {1542} (\bibinfo {year}
  {1995})}\BibitemShut {NoStop}%
\bibitem [{\citenamefont {Greiner}\ \emph {et~al.}(2002)\citenamefont
  {Greiner}, \citenamefont {Mandel}, \citenamefont {Esslinger}, \citenamefont
  {Hansch},\ and\ \citenamefont {Bloch}}]{Greiner2002}%
  \BibitemOpen
  \bibfield  {author} {\bibinfo {author} {\bibfnamefont {M.}~\bibnamefont
  {Greiner}}, \bibinfo {author} {\bibfnamefont {O.}~\bibnamefont {Mandel}},
  \bibinfo {author} {\bibfnamefont {T.}~\bibnamefont {Esslinger}}, \bibinfo
  {author} {\bibfnamefont {T.~W.}\ \bibnamefont {Hansch}}, \ and\ \bibinfo
  {author} {\bibfnamefont {I.}~\bibnamefont {Bloch}},\ }\href
  {http://dx.doi.org/10.1038/415039a} {\bibfield  {journal} {\bibinfo
  {journal} {Nature}\ }\textbf {\bibinfo {volume} {415}},\ \bibinfo {pages}
  {39} (\bibinfo {year} {2002})}\BibitemShut {NoStop}%
\bibitem [{\citenamefont {Bergquist}\ \emph {et~al.}(1988)\citenamefont
  {Bergquist}, \citenamefont {Diedrich}, \citenamefont {Itano},\ and\
  \citenamefont {Wineland}}]{Bergquist1988}%
  \BibitemOpen
  \bibfield  {author} {\bibinfo {author} {\bibfnamefont {J.}~\bibnamefont
  {Bergquist}}, \bibinfo {author} {\bibfnamefont {F.}~\bibnamefont {Diedrich}},
  \bibinfo {author} {\bibfnamefont {W.~M.}\ \bibnamefont {Itano}}, \ and\
  \bibinfo {author} {\bibfnamefont {D.}~\bibnamefont {Wineland}},\ }in\
  \href@noop {} {\emph {\bibinfo {booktitle} {Proc. 4th Symposium on Frequency
  Standards and Metrology}}},\ \bibinfo {editor} {edited by\ \bibinfo {editor}
  {\bibfnamefont {A.}~\bibnamefont {Demarchi}}}\ (\bibinfo  {publisher}
  {Springer Verlag, Heidelberg},\ \bibinfo {address} {Ancona, Italy},\ \bibinfo
  {year} {1988})\BibitemShut {NoStop}%
\bibitem [{\citenamefont {Li}\ \emph {et~al.}(1991)\citenamefont {Li},
  \citenamefont {Poggiani}, \citenamefont {Testera},\ and\ \citenamefont
  {Werth}}]{springerlink:10.1007/BF01438559}%
  \BibitemOpen
  \bibfield  {author} {\bibinfo {author} {\bibfnamefont {G.}~\bibnamefont
  {Li}}, \bibinfo {author} {\bibfnamefont {R.}~\bibnamefont {Poggiani}},
  \bibinfo {author} {\bibfnamefont {G.}~\bibnamefont {Testera}}, \ and\
  \bibinfo {author} {\bibfnamefont {G.}~\bibnamefont {Werth}},\ }\href
  {http://dx.doi.org/10.1007/BF01438559} {\bibfield  {journal} {\bibinfo
  {journal} {Zeitschrift für Physik D Atoms, Molecules and Clusters}\ }\textbf
  {\bibinfo {volume} {22}},\ \bibinfo {pages} {375} (\bibinfo {year} {1991})},\
  \bibinfo {note} {10.1007/BF01438559}\BibitemShut {NoStop}%
\bibitem [{\citenamefont {Gabrielse}\ \emph {et~al.}(2011)\citenamefont
  {Gabrielse}, \citenamefont {Kolthammer}, \citenamefont {McConnell},
  \citenamefont {Richerme}, \citenamefont {Kalra}, \citenamefont {Novitski},
  \citenamefont {Grzonka}, \citenamefont {Oelert}, \citenamefont {Sefzick},
  \citenamefont {Zielinski}, \citenamefont {Fitzakerley}, \citenamefont
  {George}, \citenamefont {Hessels}, \citenamefont {Storry}, \citenamefont
  {Weel}, \citenamefont {M\"ullers},\ and\ \citenamefont
  {Walz}}]{PhysRevLett.106.073002}%
  \BibitemOpen
  \bibfield  {author} {\bibinfo {author} {\bibfnamefont {G.}~\bibnamefont
  {Gabrielse}}, \bibinfo {author} {\bibfnamefont {W.~S.}\ \bibnamefont
  {Kolthammer}}, \bibinfo {author} {\bibfnamefont {R.}~\bibnamefont
  {McConnell}}, \bibinfo {author} {\bibfnamefont {P.}~\bibnamefont {Richerme}},
  \bibinfo {author} {\bibfnamefont {R.}~\bibnamefont {Kalra}}, \bibinfo
  {author} {\bibfnamefont {E.}~\bibnamefont {Novitski}}, \bibinfo {author}
  {\bibfnamefont {D.}~\bibnamefont {Grzonka}}, \bibinfo {author} {\bibfnamefont
  {W.}~\bibnamefont {Oelert}}, \bibinfo {author} {\bibfnamefont
  {T.}~\bibnamefont {Sefzick}}, \bibinfo {author} {\bibfnamefont
  {M.}~\bibnamefont {Zielinski}}, \bibinfo {author} {\bibfnamefont
  {D.}~\bibnamefont {Fitzakerley}}, \bibinfo {author} {\bibfnamefont {M.~C.}\
  \bibnamefont {George}}, \bibinfo {author} {\bibfnamefont {E.~A.}\
  \bibnamefont {Hessels}}, \bibinfo {author} {\bibfnamefont {C.~H.}\
  \bibnamefont {Storry}}, \bibinfo {author} {\bibfnamefont {M.}~\bibnamefont
  {Weel}}, \bibinfo {author} {\bibfnamefont {A.}~\bibnamefont {M\"ullers}}, \
  and\ \bibinfo {author} {\bibfnamefont {J.}~\bibnamefont {Walz}} (\bibinfo
  {collaboration} {ATRAP Collaboration}),\ }\href {\doibase
  10.1103/PhysRevLett.106.073002} {\bibfield  {journal} {\bibinfo  {journal}
  {Phys. Rev. Lett.}\ }\textbf {\bibinfo {volume} {106}},\ \bibinfo {pages}
  {073002} (\bibinfo {year} {2011})}\BibitemShut {NoStop}%
\bibitem [{\citenamefont {Blakestad}\ \emph {et~al.}(2009)\citenamefont
  {Blakestad}, \citenamefont {Ospelkaus}, \citenamefont {VanDevender},
  \citenamefont {Amini}, \citenamefont {Britton}, \citenamefont {Leibfried},\
  and\ \citenamefont {Wineland}}]{PhysRevLett.102.153002}%
  \BibitemOpen
  \bibfield  {author} {\bibinfo {author} {\bibfnamefont {R.~B.}\ \bibnamefont
  {Blakestad}}, \bibinfo {author} {\bibfnamefont {C.}~\bibnamefont
  {Ospelkaus}}, \bibinfo {author} {\bibfnamefont {A.~P.}\ \bibnamefont
  {VanDevender}}, \bibinfo {author} {\bibfnamefont {J.~M.}\ \bibnamefont
  {Amini}}, \bibinfo {author} {\bibfnamefont {J.}~\bibnamefont {Britton}},
  \bibinfo {author} {\bibfnamefont {D.}~\bibnamefont {Leibfried}}, \ and\
  \bibinfo {author} {\bibfnamefont {D.~J.}\ \bibnamefont {Wineland}},\ }\href
  {\doibase 10.1103/PhysRevLett.102.153002} {\bibfield  {journal} {\bibinfo
  {journal} {Phys. Rev. Lett.}\ }\textbf {\bibinfo {volume} {102}},\ \bibinfo
  {pages} {153002} (\bibinfo {year} {2009})}\BibitemShut {NoStop}%
\bibitem [{\citenamefont {Blakestad}\ \emph {et~al.}(2011)\citenamefont
  {Blakestad}, \citenamefont {Ospelkaus}, \citenamefont {VanDevender},
  \citenamefont {Wesenberg}, \citenamefont {Biercuk}, \citenamefont
  {Leibfried},\ and\ \citenamefont {Wineland}}]{PhysRevA.84.032314}%
  \BibitemOpen
  \bibfield  {author} {\bibinfo {author} {\bibfnamefont {R.~B.}\ \bibnamefont
  {Blakestad}}, \bibinfo {author} {\bibfnamefont {C.}~\bibnamefont
  {Ospelkaus}}, \bibinfo {author} {\bibfnamefont {A.~P.}\ \bibnamefont
  {VanDevender}}, \bibinfo {author} {\bibfnamefont {J.~H.}\ \bibnamefont
  {Wesenberg}}, \bibinfo {author} {\bibfnamefont {M.~J.}\ \bibnamefont
  {Biercuk}}, \bibinfo {author} {\bibfnamefont {D.}~\bibnamefont {Leibfried}},
  \ and\ \bibinfo {author} {\bibfnamefont {D.~J.}\ \bibnamefont {Wineland}},\
  }\href {\doibase 10.1103/PhysRevA.84.032314} {\bibfield  {journal} {\bibinfo
  {journal} {Phys. Rev. A}\ }\textbf {\bibinfo {volume} {84}},\ \bibinfo
  {pages} {032314} (\bibinfo {year} {2011})}\BibitemShut {NoStop}%
\bibitem [{\citenamefont {Barrett}\ \emph {et~al.}(2004)\citenamefont
  {Barrett}, \citenamefont {Chiaverini}, \citenamefont {Schaetz}, \citenamefont
  {Britton}, \citenamefont {Itano}, \citenamefont {Jost}, \citenamefont
  {Knill}, \citenamefont {Langer}, \citenamefont {Leibfried}, \citenamefont
  {Ozeri},\ and\ \citenamefont {Wineland}}]{Barrett2004}%
  \BibitemOpen
  \bibfield  {author} {\bibinfo {author} {\bibfnamefont {M.~D.}\ \bibnamefont
  {Barrett}}, \bibinfo {author} {\bibfnamefont {J.}~\bibnamefont {Chiaverini}},
  \bibinfo {author} {\bibfnamefont {T.}~\bibnamefont {Schaetz}}, \bibinfo
  {author} {\bibfnamefont {J.}~\bibnamefont {Britton}}, \bibinfo {author}
  {\bibfnamefont {W.~M.}\ \bibnamefont {Itano}}, \bibinfo {author}
  {\bibfnamefont {J.~D.}\ \bibnamefont {Jost}}, \bibinfo {author}
  {\bibfnamefont {E.}~\bibnamefont {Knill}}, \bibinfo {author} {\bibfnamefont
  {C.}~\bibnamefont {Langer}}, \bibinfo {author} {\bibfnamefont
  {D.}~\bibnamefont {Leibfried}}, \bibinfo {author} {\bibfnamefont
  {R.}~\bibnamefont {Ozeri}}, \ and\ \bibinfo {author} {\bibfnamefont {D.~J.}\
  \bibnamefont {Wineland}},\ }\href {http://dx.doi.org/10.1038/nature02608}
  {\bibfield  {journal} {\bibinfo  {journal} {Nature}\ }\textbf {\bibinfo
  {volume} {429}},\ \bibinfo {pages} {737} (\bibinfo {year}
  {2004})}\BibitemShut {NoStop}%
\bibitem [{\citenamefont {Mintert}\ and\ \citenamefont
  {Wunderlich}(2001)}]{PhysRevLett.87.257904}%
  \BibitemOpen
  \bibfield  {author} {\bibinfo {author} {\bibfnamefont {F.}~\bibnamefont
  {Mintert}}\ and\ \bibinfo {author} {\bibfnamefont {C.}~\bibnamefont
  {Wunderlich}},\ }\href {\doibase 10.1103/PhysRevLett.87.257904} {\bibfield
  {journal} {\bibinfo  {journal} {Phys. Rev. Lett.}\ }\textbf {\bibinfo
  {volume} {87}},\ \bibinfo {pages} {257904} (\bibinfo {year}
  {2001})}\BibitemShut {NoStop}%
\bibitem [{\citenamefont {Ospelkaus}\ \emph {et~al.}(2008)\citenamefont
  {Ospelkaus}, \citenamefont {Langer}, \citenamefont {Amini}, \citenamefont
  {Brown}, \citenamefont {Leibfried},\ and\ \citenamefont
  {Wineland}}]{PhysRevLett.101.090502}%
  \BibitemOpen
  \bibfield  {author} {\bibinfo {author} {\bibfnamefont {C.}~\bibnamefont
  {Ospelkaus}}, \bibinfo {author} {\bibfnamefont {C.~E.}\ \bibnamefont
  {Langer}}, \bibinfo {author} {\bibfnamefont {J.~M.}\ \bibnamefont {Amini}},
  \bibinfo {author} {\bibfnamefont {K.~R.}\ \bibnamefont {Brown}}, \bibinfo
  {author} {\bibfnamefont {D.}~\bibnamefont {Leibfried}}, \ and\ \bibinfo
  {author} {\bibfnamefont {D.~J.}\ \bibnamefont {Wineland}},\ }\href {\doibase
  10.1103/PhysRevLett.101.090502} {\bibfield  {journal} {\bibinfo  {journal}
  {Phys. Rev. Lett.}\ }\textbf {\bibinfo {volume} {101}},\ \bibinfo {pages}
  {090502} (\bibinfo {year} {2008})}\BibitemShut {NoStop}%
\bibitem [{\citenamefont {Schmidt}\ \emph {et~al.}(2006)\citenamefont
  {Schmidt}, \citenamefont {Rosenband}, \citenamefont {Koelemeij},
  \citenamefont {Hume}, \citenamefont {Itano}, \citenamefont {Bergquist},\ and\
  \citenamefont {Wineland}}]{schmidt:305}%
  \BibitemOpen
  \bibfield  {author} {\bibinfo {author} {\bibfnamefont {P.~O.}\ \bibnamefont
  {Schmidt}}, \bibinfo {author} {\bibfnamefont {T.}~\bibnamefont {Rosenband}},
  \bibinfo {author} {\bibfnamefont {J.~C.~J.}\ \bibnamefont {Koelemeij}},
  \bibinfo {author} {\bibfnamefont {D.~B.}\ \bibnamefont {Hume}}, \bibinfo
  {author} {\bibfnamefont {W.~M.}\ \bibnamefont {Itano}}, \bibinfo {author}
  {\bibfnamefont {J.~C.}\ \bibnamefont {Bergquist}}, \ and\ \bibinfo {author}
  {\bibfnamefont {D.~J.}\ \bibnamefont {Wineland}},\ }\href {\doibase
  10.1063/1.2387937} {\bibfield  {journal} {\bibinfo  {journal} {AIP Conference
  Proceedings}\ }\textbf {\bibinfo {volume} {862}},\ \bibinfo {pages} {305}
  (\bibinfo {year} {2006})}\BibitemShut {NoStop}%
\bibitem [{\citenamefont {Vogelius}\ \emph {et~al.}(2006)\citenamefont
  {Vogelius}, \citenamefont {Madsen},\ and\ \citenamefont
  {Drewsen}}]{Vogelius:Probabilistic}%
  \BibitemOpen
  \bibfield  {author} {\bibinfo {author} {\bibfnamefont {I.~S.}\ \bibnamefont
  {Vogelius}}, \bibinfo {author} {\bibfnamefont {L.~B.}\ \bibnamefont
  {Madsen}}, \ and\ \bibinfo {author} {\bibfnamefont {M.}~\bibnamefont
  {Drewsen}},\ }\href {\doibase http://dx.doi.org/10.1088/0953-4075/39/19/S31}
  {\bibfield  {journal} {\bibinfo  {journal} {Journal of Physics B: Atomic,
  Molecular and Optical Physics}\ }\textbf {\bibinfo {volume} {39}},\ \bibinfo
  {pages} {S1259} (\bibinfo {year} {2006})}\BibitemShut {NoStop}%
\bibitem [{\citenamefont {Schiller}(2007)}]{PhysRevLett.98.180801}%
  \BibitemOpen
  \bibfield  {author} {\bibinfo {author} {\bibfnamefont {S.}~\bibnamefont
  {Schiller}},\ }\href {\doibase 10.1103/PhysRevLett.98.180801} {\bibfield
  {journal} {\bibinfo  {journal} {Phys. Rev. Lett.}\ }\textbf {\bibinfo
  {volume} {98}},\ \bibinfo {pages} {180801} (\bibinfo {year}
  {2007})}\BibitemShut {NoStop}%
\bibitem [{\citenamefont {Rosenband}\ \emph {et~al.}(2008)\citenamefont
  {Rosenband}, \citenamefont {Hume}, \citenamefont {Schmidt}, \citenamefont
  {Chou}, \citenamefont {Brusch}, \citenamefont {Lorini}, \citenamefont
  {Oskay}, \citenamefont {Drullinger}, \citenamefont {Fortier}, \citenamefont
  {Stalnaker}, \citenamefont {Diddams}, \citenamefont {Swann}, \citenamefont
  {Newbury}, \citenamefont {Itano}, \citenamefont {Wineland},\ and\
  \citenamefont {Bergquist}}]{Rosenband28032008}%
  \BibitemOpen
  \bibfield  {author} {\bibinfo {author} {\bibfnamefont {T.}~\bibnamefont
  {Rosenband}}, \bibinfo {author} {\bibfnamefont {D.~B.}\ \bibnamefont {Hume}},
  \bibinfo {author} {\bibfnamefont {P.~O.}\ \bibnamefont {Schmidt}}, \bibinfo
  {author} {\bibfnamefont {C.~W.}\ \bibnamefont {Chou}}, \bibinfo {author}
  {\bibfnamefont {A.}~\bibnamefont {Brusch}}, \bibinfo {author} {\bibfnamefont
  {L.}~\bibnamefont {Lorini}}, \bibinfo {author} {\bibfnamefont {W.~H.}\
  \bibnamefont {Oskay}}, \bibinfo {author} {\bibfnamefont {R.~E.}\ \bibnamefont
  {Drullinger}}, \bibinfo {author} {\bibfnamefont {T.~M.}\ \bibnamefont
  {Fortier}}, \bibinfo {author} {\bibfnamefont {J.~E.}\ \bibnamefont
  {Stalnaker}}, \bibinfo {author} {\bibfnamefont {S.~A.}\ \bibnamefont
  {Diddams}}, \bibinfo {author} {\bibfnamefont {W.~C.}\ \bibnamefont {Swann}},
  \bibinfo {author} {\bibfnamefont {N.~R.}\ \bibnamefont {Newbury}}, \bibinfo
  {author} {\bibfnamefont {W.~M.}\ \bibnamefont {Itano}}, \bibinfo {author}
  {\bibfnamefont {D.~J.}\ \bibnamefont {Wineland}}, \ and\ \bibinfo {author}
  {\bibfnamefont {J.~C.}\ \bibnamefont {Bergquist}},\ }\href {\doibase
  10.1126/science.1154622} {\bibfield  {journal} {\bibinfo  {journal}
  {Science}\ }\textbf {\bibinfo {volume} {319}},\ \bibinfo {pages} {1808}
  (\bibinfo {year} {2008})}\BibitemShut {NoStop}%
\bibitem [{\citenamefont {Schiller}\ and\ \citenamefont
  {Korobov}(2005)}]{PhysRevA.71.032505}%
  \BibitemOpen
  \bibfield  {author} {\bibinfo {author} {\bibfnamefont {S.}~\bibnamefont
  {Schiller}}\ and\ \bibinfo {author} {\bibfnamefont {V.}~\bibnamefont
  {Korobov}},\ }\href {\doibase 10.1103/PhysRevA.71.032505} {\bibfield
  {journal} {\bibinfo  {journal} {Phys. Rev. A}\ }\textbf {\bibinfo {volume}
  {71}},\ \bibinfo {pages} {032505} (\bibinfo {year} {2005})}\BibitemShut
  {NoStop}%
\bibitem [{\citenamefont {Flambaum}\ and\ \citenamefont
  {Kozlov}(2007)}]{PhysRevLett.99.150801}%
  \BibitemOpen
  \bibfield  {author} {\bibinfo {author} {\bibfnamefont {V.~V.}\ \bibnamefont
  {Flambaum}}\ and\ \bibinfo {author} {\bibfnamefont {M.~G.}\ \bibnamefont
  {Kozlov}},\ }\href {\doibase 10.1103/PhysRevLett.99.150801} {\bibfield
  {journal} {\bibinfo  {journal} {Phys. Rev. Lett.}\ }\textbf {\bibinfo
  {volume} {99}},\ \bibinfo {pages} {150801} (\bibinfo {year}
  {2007})}\BibitemShut {NoStop}%
\bibitem [{\citenamefont {Willitsch}\ \emph {et~al.}(2008)\citenamefont
  {Willitsch}, \citenamefont {Bell}, \citenamefont {Gingell},\ and\
  \citenamefont {Softley}}]{B813408C}%
  \BibitemOpen
  \bibfield  {author} {\bibinfo {author} {\bibfnamefont {S.}~\bibnamefont
  {Willitsch}}, \bibinfo {author} {\bibfnamefont {M.~T.}\ \bibnamefont {Bell}},
  \bibinfo {author} {\bibfnamefont {A.~D.}\ \bibnamefont {Gingell}}, \ and\
  \bibinfo {author} {\bibfnamefont {T.~P.}\ \bibnamefont {Softley}},\ }\href
  {\doibase 10.1039/B813408C} {\bibfield  {journal} {\bibinfo  {journal} {Phys.
  Chem. Chem. Phys.}\ }\textbf {\bibinfo {volume} {10}},\ \bibinfo {pages}
  {7200} (\bibinfo {year} {2008})}\BibitemShut {NoStop}%
\bibitem [{\citenamefont {Carr}\ \emph {et~al.}(2009)\citenamefont {Carr},
  \citenamefont {DeMille}, \citenamefont {Krems},\ and\ \citenamefont
  {Ye}}]{1367-2630-11-5-055049}%
  \BibitemOpen
  \bibfield  {author} {\bibinfo {author} {\bibfnamefont {L.~D.}\ \bibnamefont
  {Carr}}, \bibinfo {author} {\bibfnamefont {D.}~\bibnamefont {DeMille}},
  \bibinfo {author} {\bibfnamefont {R.~V.}\ \bibnamefont {Krems}}, \ and\
  \bibinfo {author} {\bibfnamefont {J.}~\bibnamefont {Ye}},\ }\href
  {http://stacks.iop.org/1367-2630/11/i=5/a=055049} {\bibfield  {journal}
  {\bibinfo  {journal} {New Journal of Physics}\ }\textbf {\bibinfo {volume}
  {11}},\ \bibinfo {pages} {055049} (\bibinfo {year} {2009})}\BibitemShut
  {NoStop}%
\bibitem [{\citenamefont {Schmidt}\ \emph {et~al.}(2005)\citenamefont
  {Schmidt}, \citenamefont {Rosenband}, \citenamefont {Langer}, \citenamefont
  {Itano}, \citenamefont {Bergquist},\ and\ \citenamefont
  {Wineland}}]{Schmidt29072005}%
  \BibitemOpen
  \bibfield  {author} {\bibinfo {author} {\bibfnamefont {P.~O.}\ \bibnamefont
  {Schmidt}}, \bibinfo {author} {\bibfnamefont {T.}~\bibnamefont {Rosenband}},
  \bibinfo {author} {\bibfnamefont {C.}~\bibnamefont {Langer}}, \bibinfo
  {author} {\bibfnamefont {W.~M.}\ \bibnamefont {Itano}}, \bibinfo {author}
  {\bibfnamefont {J.~C.}\ \bibnamefont {Bergquist}}, \ and\ \bibinfo {author}
  {\bibfnamefont {D.~J.}\ \bibnamefont {Wineland}},\ }\href {\doibase
  10.1126/science.1114375} {\bibfield  {journal} {\bibinfo  {journal}
  {Science}\ }\textbf {\bibinfo {volume} {309}},\ \bibinfo {pages} {749}
  (\bibinfo {year} {2005})}\BibitemShut {NoStop}%
\bibitem [{\citenamefont {C\^ot\'e}\ \emph {et~al.}(2002)\citenamefont
  {C\^ot\'e}, \citenamefont {Kharchenko},\ and\ \citenamefont
  {Lukin}}]{PhysRevLett.89.093001}%
  \BibitemOpen
  \bibfield  {author} {\bibinfo {author} {\bibfnamefont {R.}~\bibnamefont
  {C\^ot\'e}}, \bibinfo {author} {\bibfnamefont {V.}~\bibnamefont
  {Kharchenko}}, \ and\ \bibinfo {author} {\bibfnamefont {M.~D.}\ \bibnamefont
  {Lukin}},\ }\href {\doibase 10.1103/PhysRevLett.89.093001} {\bibfield
  {journal} {\bibinfo  {journal} {Phys. Rev. Lett.}\ }\textbf {\bibinfo
  {volume} {89}},\ \bibinfo {pages} {093001} (\bibinfo {year}
  {2002})}\BibitemShut {NoStop}%
\bibitem [{\citenamefont {Idziaszek}\ \emph {et~al.}(2009)\citenamefont
  {Idziaszek}, \citenamefont {Calarco}, \citenamefont {Julienne},\ and\
  \citenamefont {Simoni}}]{PhysRevA.79.010702}%
  \BibitemOpen
  \bibfield  {author} {\bibinfo {author} {\bibfnamefont {Z.}~\bibnamefont
  {Idziaszek}}, \bibinfo {author} {\bibfnamefont {T.}~\bibnamefont {Calarco}},
  \bibinfo {author} {\bibfnamefont {P.~S.}\ \bibnamefont {Julienne}}, \ and\
  \bibinfo {author} {\bibfnamefont {A.}~\bibnamefont {Simoni}},\ }\href
  {\doibase 10.1103/PhysRevA.79.010702} {\bibfield  {journal} {\bibinfo
  {journal} {Phys. Rev. A}\ }\textbf {\bibinfo {volume} {79}},\ \bibinfo
  {pages} {010702} (\bibinfo {year} {2009})}\BibitemShut {NoStop}%
\bibitem [{\citenamefont {Poulsen}\ \emph {et~al.}(2012)\citenamefont
  {Poulsen}, \citenamefont {Miroshnychenko},\ and\ \citenamefont
  {Drewsen}}]{Poulsen:SBC}%
  \BibitemOpen
  \bibfield  {author} {\bibinfo {author} {\bibfnamefont {G.}~\bibnamefont
  {Poulsen}}, \bibinfo {author} {\bibfnamefont {Y.}~\bibnamefont
  {Miroshnychenko}}, \ and\ \bibinfo {author} {\bibfnamefont {M.}~\bibnamefont
  {Drewsen}},\ }\href@noop {} {\enquote {\bibinfo {title} {Efficient ground
  state cooling of an ion in a large room temperature linear paul trap with a
  sub-hertz heating rate},}\ } (\bibinfo {year} {2012}),\ \bibinfo {note}
  {accepted PRA Rapid Comm. (arXiv:1205.1949).}\BibitemShut {Stop}%
\bibitem [{\citenamefont {Roos}\ \emph {et~al.}(1999)\citenamefont {Roos},
  \citenamefont {Zeiger}, \citenamefont {Rohde}, \citenamefont {N\"agerl},
  \citenamefont {Eschner}, \citenamefont {Leibfried}, \citenamefont
  {Schmidt-Kaler},\ and\ \citenamefont {Blatt}}]{PhysRevLett.83.4713}%
  \BibitemOpen
  \bibfield  {author} {\bibinfo {author} {\bibfnamefont {C.}~\bibnamefont
  {Roos}}, \bibinfo {author} {\bibfnamefont {T.}~\bibnamefont {Zeiger}},
  \bibinfo {author} {\bibfnamefont {H.}~\bibnamefont {Rohde}}, \bibinfo
  {author} {\bibfnamefont {H.~C.}\ \bibnamefont {N\"agerl}}, \bibinfo {author}
  {\bibfnamefont {J.}~\bibnamefont {Eschner}}, \bibinfo {author} {\bibfnamefont
  {D.}~\bibnamefont {Leibfried}}, \bibinfo {author} {\bibfnamefont
  {F.}~\bibnamefont {Schmidt-Kaler}}, \ and\ \bibinfo {author} {\bibfnamefont
  {R.}~\bibnamefont {Blatt}},\ }\href {\doibase 10.1103/PhysRevLett.83.4713}
  {\bibfield  {journal} {\bibinfo  {journal} {Phys. Rev. Lett.}\ }\textbf
  {\bibinfo {volume} {83}},\ \bibinfo {pages} {4713} (\bibinfo {year}
  {1999})}\BibitemShut {NoStop}%
\end{thebibliography}
%

\end{document}